\documentclass{moriond}

\def\be{\begin{equation}}
\def\ee{\end{equation}}
\def\bea{\begin{eqnarray}}
\def\eea{\end{eqnarray}}

\newcommand{\Photo}{\includegraphics[height=35mm]{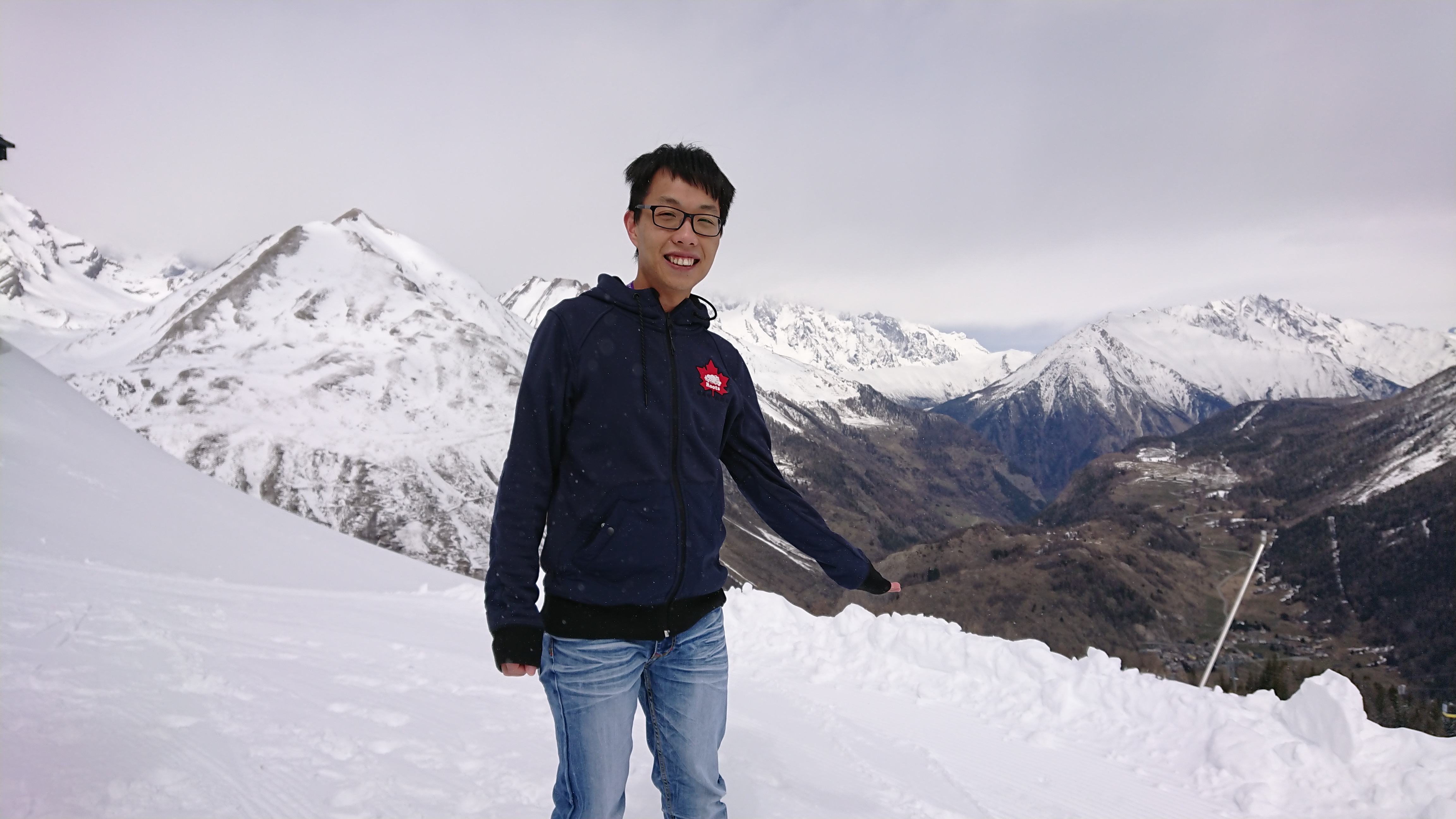}}

\begin{document}
\vspace*{4cm}
\title{The Hierarchy Problem and the Top Yukawa}

\author{ Andreas Bally, Yi Chung \footnote{Speaker}, and Florian Goertz}

\address{Max-Planck-Institut f\"ur Kernphysik, Saupfercheckweg 1,\\ 69117 Heidelberg, Germany}

\maketitle\abstracts{
In this talk, an alternative to top partner solutions and its consequences on phenomenology are discussed. The hierarchy problem from the top loop contribution is solved by mitigating the top Yukawa coupling at high scales. In this scenario, the new degrees of freedom appearing at the cut-off scale of the top loop should then be some new top-philic particles instead of traditional top partners. The idea can be directly tested through measurements in top physics, including $t\bar{t}h$, $t\bar{t}$ differential cross section, and $t\bar{t}t\bar{t}$ cross section.}

\section{Introduction}

The hierarchy problem is one of the long-standing problems in the Standard Model (SM) of particle physics. The problem is growing severe after the discovery of a light Higgs boson without any other new degrees of freedom. The Higgs boson's mass term receives quadratically divergent contributions from loops of all the particles it interacts with. Among all the loops, the top quark loop gives the largest contribution as
\begin{equation}\label{toploop}
\Delta m_H^2|_{\rm top}
\sim-i\,2N_c\,y_t^2 \int_{}^{} \frac{d^4k}{(2\pi)^4}\frac{k^2+m_t^2}{(k^2-m_t^2)^2}= -\frac{3}{8\pi^2}y_t^2
\left[\Lambda_{\rm NP}^2-3\,m_t^2\,{\rm ln}\left(\frac{\Lambda^2_{\rm NP}}{m_t^2}\right)+\cdots\right]~,
\end{equation}
where we only kept the $\Lambda_{\rm NP}$-dependent terms.

Traditionally, top partners are introduced to solve the hierarchy problem. The new contribution $\Delta m_H^2|_{\rm top\;partner}$ from the top partner loop cancels the quadratically divergent term $\Lambda_{\rm NP}^2$, which is guaranteed by some new symmetry, such as supersymmetry. However, the symmetry can not be exact and thus the Higgs mass term is still generated due to the difference between the top quark mass and the top partner mass as
\begin{equation}\label{toploop+}
\Delta m_H^2|_{\rm top}+\Delta m_H^2|_{\rm top \; partner}
\sim -\frac{3}{8\pi^2}y_t^2 M_T^2\,{\rm ln}\left(\frac{\Lambda_{\rm NP}^2}{M_T^2}\right)~,
\end{equation}
where $M_T$ is the mass of the top partner. To get the light Higgs mass, the Naturalness principle suggests top partners with $M_T\sim 500$ GeV.

However, after years of searches by the LHC, the bounds on the mass of colored top partners have reached around $1500$ GeV for both scalar top partners \cite{CMS:2020pyk,ATLAS:2020aci} and fermionic top partners \cite{CMS:2019eqb,ATLAS:2018ziw,ATLAS:2022ozf}. The absence of top partners starts challenging the naturalness of these traditional models due to the required fine-tuning. To reduce fine-tuning, one alternative is to introduce uncolored top partners. Since they are harder to be produced, the bound on their mass is rather weak. It would be even better if the top partner is a SM singlet, which is known as Neutral Naturalness, like in Twin Higgs models \cite{Chacko:2005pe}. However, this alternative is still based on the idea of symmetry and the cancellation between $\Delta m_H^2|_{\rm top}$ and $\Delta m_H^2|_{\rm top \; partner}$. In this work \cite{Bally:2022naz}, we explore another direction that does not require top partners. The idea is to have a strong dependence of the top Yukawa coupling on the energy scale $y_t=y_t(k^2)$, which can make the top loop contribution converge. For example, taking
\begin{equation}\label{toploopnew}
y_t(k^2)=y_t\left(\frac{\Lambda_T^2}{-k^2+\Lambda_T^2}\right), \ \ 
\Delta m_H^2|_{\rm top}\sim-i\,2N_c \int_{}^{} \frac{d^4k}{(2\pi)^4}\,y_t^2(k^2)\frac{k^2+m_t^2}{(k^2-m_t^2)^2}~\sim -\frac{3}{8\pi^2}y_t^2 \Lambda_T^2~,
\end{equation}
where the top loop contribution is controlled by $\Lambda_T$, which is the mass scale of new degrees of freedom responsible for the nontrivial behavior of top Yukawa coupling.

\section{Zoom in the Top Yukawa vertex}

\begin{figure}[!h]
\begin{minipage}{1.0\linewidth}
\centerline{\includegraphics[width=1\linewidth]{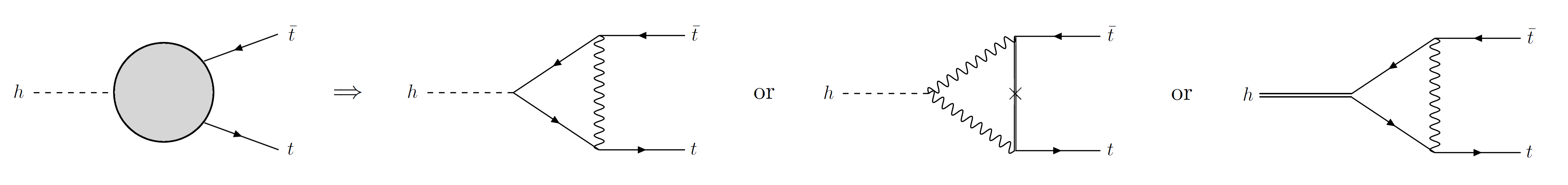}}
\end{minipage}
\caption[]{Three possible UV completions for the nontrivial behavior of the top Yukawa coupling.}
\label{zoomin}
\end{figure}

To realize the idea, the top Yukawa coupling must come from a more intricate origin, which implies new physics in the top Yukawa vertex at high scales. There are several possibilities as shown in Fig. \ref{zoomin} and each case (from left to right) is described as follows (for more details, see our work \cite{Bally:2022naz}).

The first possibility corresponds to  \textbf{“large $y_t$ running”}, where an additional top-philic interaction introduces large running to reduce the top Yukawa coupling at high scales. Negative contributions to the $y_t$ running already exist in the SM but are not large enough. Therefore, we need a stronger interaction to bring down the top Yukawa coupling at the desired scale. The wavy propagator can be either a vector boson or a scalar boson. The only requirement is that the coupling needs to be strong enough to have an impact on the top Yukawa coupling.

Second,  \textbf{``radiative $y_t$ generation''}, where the top Yukawa coupling is not present at the tree level but is only generated at the one-loop level. In this scenario, new top-philic bosons (wavy line) and vector-like fermions (double line) are introduced to connect the Higgs boson and the top quarks. The top Yukawa is only obtained after integrating out these new degrees of freedom, which also means, above the mass scale of these intermediate states, the top Yukawa will drop as desired. The interaction among these new particles should also be strong enough to reproduce the correct value of the top Yukawa coupling.

The last one \textbf{``$y_t$ from four-fermion interactions''}, different from the first two scenarios, is only valid for composite Higgs models. In the first two scenarios, an elementary Higgs and top quark are assumed so the modification can only happen at the one-loop level, which thus requires strong couplings. However, if the Higgs is composite, the top Yukawa coupling can arise from four-fermion interactions. The idea can be traced back to Extended Technicolor models \cite{Dimopoulos:1979es,Eichten:1979ah} but now evolved into a modern version with a light composite Higgs, usually called Extended Hypercolor \cite{Ferretti:2013kya,Cacciapaglia:2015yra}. In this scenario, the top Yukawa originates from a dimension-six operator, which will also decrease when passing the mass scale of the extended-hypercolor bosons.

\section{Phenomenology}

When talking about testing new physics, we always think about searches of new resonances. However, it is not straightforward in this case as the main idea is to modify the top Yukawa at high scales. Direct searches require us to first specify the new degrees of freedom responsible for the modification, which includes many possibilities as shown in the last section. Moreover, for each scenario, there are different difficulties. For the first two, the modification happens at the one-loop level, which means the quantum number of the new states is not uniquely determined. Also, the requirement of strong couplings implies some broad resonances, which makes them hard to look for. For the extended-hypercolor bosons, again, the quantum number is diverse and depends on the hypercolor group \footnote{The only exception is the top-philic $Z'$ boson which will be studied in our next work.}. Overall, it is hard to perform direct searches due to the requirements of the UV theory and strong couplings. However, the direct test of the idea can be realized in other measurements in top physics.

\subsection{Running top Yukawa and $t\bar{t}h$ differential cross section}

\begin{figure}[!h]
\begin{minipage}{1.0\linewidth}
\centerline{\includegraphics[width=0.7\linewidth]{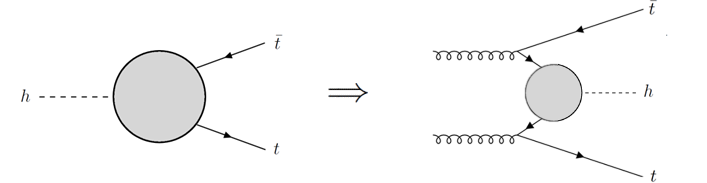}}
\end{minipage}
\caption[]{The $y_t=y_t(k^2)$ can be tested through $t\bar{t}h$ differential cross section.}
\label{tth}
\end{figure}

Starting with the direct probe of the top Yukawa coupling at high scales, the desired process is $t\bar{t}h$ final state as shown in Fig. \ref{tth}. The nontrivial behavior of the top Yukawa coupling at high scales will reveal itself in differential momentum
distributions of $t\bar{t}h$ production \cite{MammenAbraham:2021ssc,Bittar:2022wgb}. However, based on the $t\bar{t}h$ cross section, it is hard to achieve the desired sensitivity in the LHC era.

\subsection{Running top mass and $t\bar{t}$ differential cross section}

\begin{figure}[!h]
\begin{minipage}{1.0\linewidth}
\centerline{\includegraphics[width=1\linewidth]{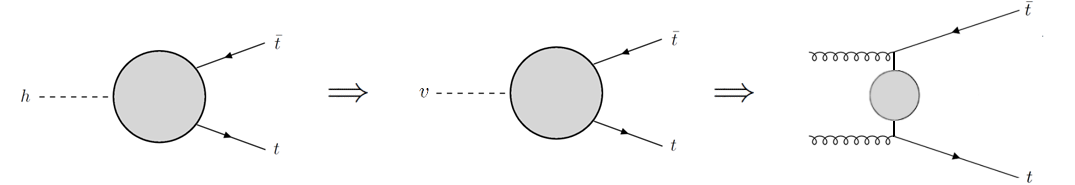}}
\end{minipage}
\caption[]{The $m_t=m_t(k^2)$ can be tested through $t\bar{t}$ differential cross section.}
\label{tt}
\end{figure}

Besides probing the top Yukawa coupling at high scales directly, we can also measure the top quark mass at high scales instead, which is like simply replacing the Higgs with its VEV. Without a Higgs, the measurement can then be done from $t\bar{t}$ final states, which has a much larger total cross section. Such measurement has already been done by the CMS collaboration using run 2 data with an integrated luminosity of 35.9 fb$^{-1}$ and the result presents the running top mass up to $1$ TeV \cite{CMS:2019jul}, which constraints the idea we present. However, the result has also been reinterpreted in another theory paper \cite{Defranchis:2022nqb} but considering the energy scale only half of the value in the original CMS paper. The result is then only sensitive up to an energy scale of $500$ GeV. In this case, the bound on the top mass running becomes weaker and light new physics with less fine-tuning is still possible.

\subsection{Top-philic interaction and $t\bar{t}t\bar{t}$ cross section}

\begin{figure}[!h]
\begin{minipage}{1.0\linewidth}
\centerline{\includegraphics[width=1\linewidth]{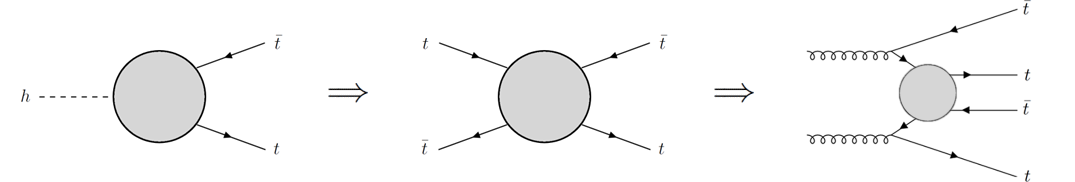}}
\end{minipage}
\caption[]{New top-philic interactions can be tested through $t\bar{t}t\bar{t}$ cross section.}
\label{tttt}
\end{figure}

Modifying top Yukawa coupling always requires some new top-philic interaction, which will unavoidably introduce additional contributions to the four top cross section. Generic estimation predicts four top operators with coefficients $\sim C/\Lambda_T^2$. To determine the Lorentz structure of the operators and their coefficients, the UV completion is again required. However, a sizable contribution is expected and it turns out the four top quarks cross section might be the most promising channels to look for. In the SM, the prediction was updated a few months ago with next-to-leading logarithmic (NLL') accuracy \cite{vanBeekveld:2022hty}. From the experimental side, both ATLAS and CMS just announced the observation \cite{ATLAS:2023ajo,CMS:2023ica} in Moriond. The latest result has started constraining the "large $y_t$ running" scenario with strongly coupled top-philic bosons. The other scenarios are expected to be confirmed or ruled out within the LHC era.

\section{Conclusion}

The top quark plays the most important role in the hierarchy problem. Traditionally, top partners are introduced to cancel the top-loop contribution. In this talk, we discuss an alternative where the top-loop contribution is reduced by modifying the running of the top Yukawa coupling. The new degrees of freedom expected to appear at the Naturalness scale $\Lambda_T\sim 500$ GeV will be new top-philic particles instead of top partners. Direct tests of the idea can be realized in top physics, including $t\bar{t}h$, $t\bar{t}$, and $t\bar{t}t\bar{t}$ production, where Naturalness might still be hidden !

\end{document}